\documentclass[12pt]{iopart}
\usepackage{amsfonts,amssymb,latexsym}
\usepackage{graphicx,epsf}
\usepackage{color}

\newcommand{\braket}[2]{\left \langle #1 | #2 \right\rangle}

\newcommand{\be}{\begin{equation}}
\newcommand{\ee}{\end{equation}}
\newcommand{\ba}{\begin{eqnarray}}
\newcommand{\ea}{\end{eqnarray}}
\newcommand{\ignore}[1]{}

\newcommand{\ket}[1]{\left | {#1} \right \rangle }

\newcommand{\bra}[1]{\left \langle {#1} \right | }

\newcommand{\sig}[2]{\sigma_{#1}^{#2}}
\newcommand{\la}{\lambda}

\usepackage{hyperref}

\begin{document}

\title{Quantum geometric tensor away from Equilibrium}

\author{Davide Rattacaso$^1$, Patrizia Vitale$^1$$^2$ and Alioscia Hamma$^3$$^4$}
\address{$^1$ Dipartimento di Fisica Ettore Pancini, Universit\`a di Napoli Federico II}
\address{$^2$ INFN-Sezione di Napoli}
\address{$^3$ Physics Department,  University of Massachusetts Boston,  02125, USA}
\address{$^4$ Univ. Grenoble Alpes, CNRS, LPMMC, 38000 Grenoble, France}

\begin{abstract}
The manifold of ground states of a family of quantum Hamiltonians can be endowed with a quantum geometric tensor whose singularities signal quantum phase transitions and give a general way to define quantum phases. In this paper, we show that the same information-theoretic and geometrical approach can be used to describe the geometry of quantum states away from equilibrium. We construct the quantum geometric tensor $Q_{\mu\nu}$ for ensembles of states that evolve in time and study its phase diagram and equilibration properties. If the initial ensemble is the manifold of ground states, we show that the phase diagram is conserved, that the geometric tensor equilibrates after a quantum quench, and that its time behavior is governed by out-of-time-order commutators (OTOCs). We finally demonstrate our results in the exactly solvable Cluster-XY model.
\end{abstract}
\vspace{2pc}


\section{Introduction} 
The notion of a quantum phase is that of an equivalence class of quantum states, in particular, of ground states of a family of Hamiltonians. The states in the same phase are those which  look alike according to some salient characteristics that allow for a classification in different classes. For instance, all the states that break the symmetry of a parent Hamiltonian in a certain way are labeled by the correspondent local order parameter. A quantum phase transition\cite{sachdev} is the transition between different quantum phases and is usually signalised  by the shrinking of the gap between the ground state and the first excited state. In order to go beyond symmetry breaking phases, it has proven very fruitful to use tools from quantum information theory, in particular measures of distinguishability of quantum states. The main idea is that, from a state in a quantum phase, one can find other states that are infinitesimally close and in this way it is possible to  connect any two states in the same phase. On the other hand, crossing a quantum phase transition means that at some point the distance between two quantum states is not analytic. 

A metric on the space of quantum states can be easily obtained by  the so called {\em fidelity} $\mathcal F=|\braket{\phi}{\psi}|$ between the pure quantum states $\ket{\phi},\ket{\psi}$. As Wooters showed in\cite{wooters}, the quantity $d_{FS}(\phi,\psi) =\cos^{-1} \mathcal F$ represents the maximum over all the possible projective measurements of the Fisher-Rao statistical distance between the probability distributions obtained from $\ket{\phi},\ket{\psi}$. {and therefore it estimates the distinguishability of quantum states through repeated measurements.} The infinitesimal version of this metric distance gives rise to the Fubini-Study metric $d^2_{FS}(\psi, \psi+\delta\psi)\simeq 2(1-\mathcal F)${, that encodes the distinguishability between states that are infinitesimally distant in the space of parameters}. This information-theoretic distance is closely related to the quantum geometric tensor $Q_{\mu\nu}$, namely the natural metric structure on the projective Hilbert space\cite{provost}. Quantum phase transitions can then be studied in a more elegant and general way by looking at the scaling of the norm of such tensor\cite{zanardiqgt, zanardiscaling}. It turns out that quantum critical points are marked by divergences of the norm of the real part of  $Q_{\mu\nu}$. This approach has  proven useful to study topological quantum phase transitions\cite{fidelitytop} and it has been generalised to mixed (e.g., thermal) states\cite{zanardibures}.

An important question is that of clarifying the notion of quantum phase for states away from equilibrium. This would prove very useful to understand dynamical phase transitions\cite{polkovnikov} like the transition between thermal and many-body localised states\cite{abaninrmp}, the classical-quantum transition\cite{cl-quantum1,cl-quantum2,cl-quantum3}, the onset of chaotic behavior or the transition between scrambling and unscrambling behaviour\cite{yoshida, otocANDchaos}. Quantum phase transitions have been shown\cite{zanardiLE} to affect the decay of the Loschmidt Echo, which is a way to evaluate the sensibility of the dynamics to perturbations of the system. Since quantum criticality is described by the geometry of quantum states, and it affects the behavior of the dynamics, a question best to be answered: how does this geometric structure behave for states away from equilibrium?

To start with, we consider  a manifold of ground states and move them away from equilibrium by means of a quantum quench\cite{criticalquench,colloquium}. 
Thanks to  the formalism of quantum quenches we construct   a foliated manifold of quantum states $\mathcal M_t$ and show that it    can be given a metric structure $Q_{\mu\nu}(t)$. We  show that the phase diagram on $\mathcal M_t$ is conserved, find conditions for the equilibration of the geometric tensor and find that the time evolution of the tensor may be expressed in terms of out-of time-order commutators (OTOCs). We finally apply our results  to an integrable model, the Cluster-XY model\cite{hammaclusterxy}.

\section{Setup} 
Consider a Hamiltonian $H(\lambda)$ smooth in the parameters $\lambda= (\lambda^1,..,\lambda^n)$ and consider the mapping $\lambda\mapsto \ket{\psi (\lambda)}$ to the (unique) ground state of $H(\lambda)$. The projective Hilbert space ${\mathcal PH}$ of the rays is the base manifold of a $U(1)$ fiber bundle naturally endowed with a complex metrics $\mathcal{G}(u,v) = \langle u|(1-\ket{\psi (\lambda)}\bra{\psi (\lambda)} ) v\rangle$. {This bundle structure reflects the fact that normalised vectors that differ only by a global complex phase represent the same quantum state. In order to deal with the manifold $\mathcal{M}$ of ground states we need to pull back this metric.} Such a pull-back yields the Hermitian {\em quantum geometric tensor} $Q_{\mu\nu}\equiv\langle\partial_\mu\psi_{0}|(1-|\psi_{0}\rangle\langle\psi_{0}|)|\partial_\nu\psi_{0}\rangle$. The real part of the latter, $g_{\mu\nu} = \mathfrak R Q_{\mu\nu}$,  is a Riemannian real geometric tensor on $\mathcal M$ while its imaginary part is the Berry adiabatic curvature. {The Riemannian metric encodes the distinguishability between states, while the curvature is related to the Berry phase acquired by a state when it is driven through an adiabatic cycle on the manifold $\mathcal{M}$}. For real Hamiltonians the ground state manifold is real and one has $Q_{\mu\nu}=g_{\mu\nu}$.

Now let us introduce  the unitary operator $U_t(\lambda)$ that generates time evolution on  $\mathcal M$.  This operation defines a new family of states,  $\mathcal M_t=\{|\psi_{0t}(\lambda)\rangle=U(\lambda,t)|\psi_0(\lambda)\rangle\}$. We can pull back the complex metric $\mathcal G$ to the manifold $\mathcal M_t$  and similarly obtain the time dependent quantum geometric tensor
$
Q_{\mu\nu}(t)\equiv\langle\partial_\mu\psi_{0t}|(1-|\psi_{0t}\rangle\langle\psi_{0t}|)|\partial_\nu\psi_{0t}\rangle.
$

{
An useful expression for the quantum geometric tensor on the manifold $\mathcal{M}$ has been calculated in\cite{zanardiscaling}. It reads 
$$
Q_{\mu\nu}=\sum_{n\neq0}\frac{\langle\psi_0|\partial_\mu H|\psi_n\rangle\langle\psi_n|\partial_\nu H|\psi_0\rangle}{(E_0-E_n)^2},
$$
 where we have omitted $\lambda$ for the sake of simplicity.
Since $\mathcal{M}_t$ is the manifold of (unique) groud states of the smooth family of Hamiltonians $H(\lambda)_{-t}\equiv U(\lambda,t)H(\lambda)U(\lambda,t)^\dag$, this equation holds also for $Q_{\mu\nu}(t)$ in the following form
\begin{equation}
Q_{\mu\nu}(t)=\sum_{n\neq0}\frac{\langle\psi_{0t}|\partial_\mu H_{-t}|\psi_{nt}\rangle\langle\psi_{nt}|\partial_\nu H_{-t}|\psi_{0t}\rangle}{(E_0-E_n)^2}\nonumber
\end{equation}}
Let $q_a(t)$ be the eigenvalues of eigenvector $\ket{v_a(t)}$ for the geometric tensor $Q_{\mu\nu}(t)$ in the $\mu\nu$ space. {Exploiting the time dependent coordinate map that diagonalizes the QGT, we obtain
\begin{eqnarray}
q_a(t)&=&\sum_{n\neq0}\frac{\langle\psi_{0t}|\partial_a H_{-t}|\psi_{nt}\rangle\langle\psi_{nt}|\partial_a H_{-t}|\psi_{0t}\rangle}{(E_0-E_n)^2}=\sum_{n\neq0}\frac{|\langle\psi_0|U^\dag\partial_a(UHU^\dag)U|\psi_n\rangle|^2}{(E_0-E_n)^2} \nonumber\\
&=&\sum_{n\neq0}\frac{|\langle\psi_0|\partial_aH+[H,iD_a]|\psi_n\rangle|^2}{(E_0-E_n)^2}\nonumber
\end{eqnarray}
where we have used  $\partial_a U^\dag U=-U^\dag\partial_a U$ and defined $D_a\equiv-i\partial_a U^\dag U$.
For the sake of not burdening the notation, from now on we drop the index $a$ obtaining
\begin{eqnarray}\label{gt}
q(t)&=&\sum_{n\neq0}\frac{|\langle\psi_0|\partial H+[H,iD]|\psi_n\rangle|^2}{(E_0-E_n)^2}
\end{eqnarray}
}
So far, the evolution operator $U(\lambda,t)$ is completely general. Now we specialize it to the case in which time evolution is obtained by a sudden quantum quench. To this end, we define another family of Hamiltonians $H^q(\lambda)$ on the same manifold $\mathcal M$, the so called {\em quench} Hamiltonian, and consider the mapping $\lambda\mapsto \ket{\psi_{0t}(\lambda)}= U(\lambda,t)\ket{\psi_{0}(\lambda)}$ where $U(\lambda,t)=\exp(-it H^q(\lambda))$ is the unitary evolution operator associated to $H^q(\lambda)$. Typically, a quench can be obtained by posing  $H^q(\lambda) = H(\lambda+\delta\lambda)$, where $\delta\lambda$ is a small variation of the $\lambda$ parameters on $\mathcal M$.

For a quantum quench the expression of $D$ can be considerably simplified. {
Indeed, when the time evolution $U$ is generated by the quench Hamiltonian $H^q$ the following equation holds:
\begin{eqnarray}
\partial U^\dag U&=&\lim_{d\rightarrow0}\frac1d[e^{itH^q(\lambda+d)}e^{-itH^q(\lambda)}-e^{itH^q(\lambda)}e^{-itH^q(\lambda)}]\nonumber\\
&=&\lim_{d\rightarrow0}\frac1d[e^{it(H^q+d\partial H^q)}e^{-itH^q}-1]\nonumber
\end{eqnarray}
whose time derivative  gives
\begin{eqnarray}
\frac {d}{dt}[\partial U^\dag U]&=&\frac{d}{dt}\Big[\lim_{d\rightarrow0}\frac1d[e^{it(H^q+d\partial H^q)}e^{-itH^q}-1]\Big]=\lim_{d\rightarrow0}\frac1d\frac{d}{dt}\Big[e^{it(H^q+d\partial H^q)}e^{-itH^q}-1\Big]\nonumber\\
&=&iU^\dag\partial H^qU\nonumber
\end{eqnarray}
If we integrate both the RHS and the LHS of the last equation and take into account that when $t=0$ $\partial U^\dag U=0$, we obtain that for a quantum quench $D=\int_0^tdt'U(t')^\dag\partial H^qU(t')$.
}

In the above scheme, every point $\ket{\psi_{0}(\lambda)}$ in the ground state manifold is quenched with a different Hamiltonian $H^q(\lambda)$.
A simplified protocol for the quantum quench consists in starting from a single point and using it as a seed for the evolution of the whole manifold. We obtain this by preparing the initial state in the ground state $\ket{\psi_0(\lambda_0)}$ of a fixed $H(\lambda_0)$ and then evolving with $H^q = H(\lambda)$. In this case, the term $ \partial H$ in Eq.(\ref{gt}) vanishes and we can define a simplified geometric tensor
\begin{equation}\label{simp_tens}
q_1(t)\equiv\sum_{n\neq0}{|\langle\psi_0|[H,iD]|\psi_n\rangle|^2}{(E_0-E_n)^{-2}}
\end{equation}

Notice that, since $$q(0) = \sum_{n\neq0} {|\langle\psi_0|\partial H|\psi_n\rangle|^2}{(E_0-E_n)^{-2}},$$ one can exploit the triangular inequality to bound the absolute value of the $q(t)$.
{
Indeed, it is immediate to define the complex vectors $a_n$ and $b_n(t)$ with $n\in\{1;2;...;N\}$ such that
\begin{eqnarray}
q(0)&=&\sum_{n\neq0}\frac{|\langle\psi_0|\partial H|\psi_n\rangle|^2}{(E_0-E_n)^2}=|a|^2\nonumber\\
q_1(t)&=&\sum_{n\neq0}\frac{|\langle\psi_0|[H,iD]|\psi_n\rangle|^2}{(E_0-E_n)^2}=|b(t)|^2\nonumber\\
q(t)&=&\sum_{n\neq0}\frac{|\langle\psi_0|\partial H+[H,iD]|\psi_n\rangle|^2}{(E_0-E_n)^2}=|a+b(t)|^2\nonumber
\end{eqnarray}
where $|a|\equiv\sqrt{\sum_na_n^*a_n}$ is the norm induced by the inner product.
In this way the triangular inequality for the vectors $a_n$ and $b_n(t)$ becomes the following useful bound for $q(t)$
\begin{equation}
q(0)+q_1(t)-2\sqrt{q(0)q_1(t)}\leq q(t)\leq q(0)+q_1(t)+2\sqrt{q(0)q_1(t)}
\end{equation}
}
\ignore{Moreover, defining the operator $O$  by $O|\psi_n\rangle=(E_0-E_n)^{-1}|\psi_n\rangle$ for all $n\neq0$, we obtain
\ba
q(t)&=&\sum_{n\neq0}\langle\psi_0|\partial HO+iD|\psi_n\rangle\langle\psi_n|O\partial H-iD|\psi_0\rangle\nonumber\\
&=&q(0) +q_1(t) +I(t)
\ea
 where one has defined $I(t)\equiv\sum_{n\neq0}-i\langle\psi_0|\partial HO|\psi_n\rangle\langle\psi_n|D|\psi_0\rangle+i\langle\psi_0|D|\psi_n\rangle\langle\psi_n|O\partial H|\psi_0\rangle$.
 }
 
\section{Time evolution of the Phase Diagram} 
The zero-time phase diagram is a consequence of the locality of the Hamiltonian, as divergences of the rescaled quantum geometric tensor $q_a/N$ can only happen when the gap $\Delta \equiv E_1-E_0$ with the first excited state closes\cite{zanardiscaling}. In order to show the time evolution of the phase diagram on $\mathcal M_t$, we need to exploit locality again. From now on, we will be interested in {\em local} Hamiltonians, that is, sum of local operators $H(\lambda) = \sum_i H_i(\lambda)$, and similarly for the quench Hamiltonian $H^q(\lambda)= \sum_i H^q_i(\lambda)$. 

Let us start with showing that $q_a(t)$ can be written as a connected correlation function for $D_a$. We use the simplified quench so that we can drop the term $ \partial_aH$ in Eq.(\ref{gt}). We also drop {again }the subscript $a$ for the sake of making the notation lighter.

{
If we make explicit the commutator in Eq. (\ref{simp_tens}) and act with the Hamiltonian $H$ on the state $\ket{\psi_0}$ we obtain that $q_1(t)
=\sum_{n\neq0}|\langle\psi_0|D |\psi_n\rangle|^2
=\sum_{n}|\langle\psi_0|D |\psi_n\rangle|^2-\langle\psi_0|D |\psi_0\rangle^2
=\langle\psi_0|D ^2|\psi_0\rangle-\langle\psi_0|D |\psi_0\rangle^2\equiv\langle D^2\rangle_C$.
Writing the operator $D$ explicitly and  exploiting the locality and the translational invariance of the Hamiltonian this last equation becomes
\begin{equation}\label{gcor}
q_1(t)/N=\sum_{j}\int_0^tdt'\int_0^tdt''\langle \partial H^q_0(t')\partial H^q_j(t'')\rangle_C
\end{equation}
}
We  see  that the geometric tensor is given by the sum of  unequal-time connected correlation functions  of the variation of the quench Hamiltonian. In the general quench  scheme instead, we can upper bound $|q_a(t)|\le \Delta^{-2}\langle \tilde{D}^2(H)\rangle_C$, where we have defined the covariant derivative $\tilde{D}_a(X) = \partial_a X - [iD_a,X]$. 

The locality of the Hamiltonian and its spectral properties can now be exploited to upper bound the norm of $q_1(t)$. On  generalising a result about the Lieb-Robinson bounds\cite{truncation}, it is possible to prove the following: if two local normalised operators $O_A$ and $O_B$ are separated by a distance $d_{AB}$, the unequal-time correlation functions are upper bounded by
$|\langle O_A(t')O_B(t'')\rangle|_\psi\leq k e^{-\frac{d_{AB}}{\chi+a}}(e^{\frac{v_{LR}|t'|}{\chi+a}}+e^{\frac{v_{LR}|t''|}{\chi+a}})^2$. Here,  $v_{LR}$ is the maximum speed of the interactions, the so-called Lieb-Robinson speed, $\chi$ is the correlation length of the state $\psi$, that is, the initial amount of correlations in the state,  and $a$ is the lattice spacing. Moreover,  $k$ is a constant that does not depend on the size of the system.

{
Let us prove this statement.  As a consequence of Lieb-Robinson bound, given a local normalized operator $O_A$ with support on $A$, the following inequality holds\cite{truncation}:
\begin{equation}\label{corr01}
||O_A(t)-O_A^l(t)||\leq c|A|\exp\Big(-\frac{l-v_{LR}|t|}{a/2}\Big)
\end{equation}
where $O_A^l(t)$ is the restriction of $O_A$ to the sites of the lattices that are less than $l$ away from $A$, 
$|A|$ is the cardinality of $A$, $a$ is the lattice spacing and $c$ is a constant that depends only on the maximum norm of the local interactions and on the maximum degree of the interaction vertices.
On the other hand, for the exponential clustering theorem\cite{PhysRevLett.93.140402} the following upper bound on the ground states of a gapped system holds:
\begin{equation}\label{corr02}
|\langle O_AO_B\rangle_C|\leq\alpha\exp\Big(-\frac{d_{AB}}{\chi}\Big)
\end{equation} 
where $O_A$ and $O_B$ are normalized local operators, $\chi$ is the correlation length, $d_{AB}$ is the distance between the supports $A$ and $B$ and $\alpha$ does not depend on the size of the system.
Once $\Delta_A(l',t')\equiv O_A(t')-O_A^{l'}(t')$ is defined we can write
\begin{eqnarray}
&&|\langle O_A(t')O_B(t'')\rangle_C|=|\langle\Delta_A(l',t')\Delta_B(l'',t'')\rangle_C|+|\langle\Delta_A(l',t')O_B^{l''}(t'')\rangle_C|\nonumber\\
&&+|\langle O_A^{l'}(t')\Delta_B(l'',t'')\rangle_C|+|\langle O_A^{l'}(t')O_B^{l''}(t'')\rangle_C|\nonumber
\end{eqnarray}
Since $\langle O_AO_B\rangle_C\leq\langle O_AO_B\rangle\leq||O_AO_B||\leq||O_A|||||O_B||$ and $||O_A^l||\leq||O_A||=1$ the following inequality holds:
\begin{eqnarray}
&&|\langle O_A(t')O_B(t'')\rangle_C|\nonumber\\
&&\leq||\Delta_A(l',t')||||\Delta_B(l'',t'')||+||\Delta_A(l',t')||\nonumber\\
&&+||\Delta_B(l'',t'')||+|\langle O_A^{l'}(t')O_B^{l''}(t'')\rangle_C|.\nonumber
\end{eqnarray}
Thus,  the triangular inequality $||\Delta_A(l',t')||\leq||O_A(t')||+||O_A^{l'}(t')||\leq2$ leads us to
\begin{eqnarray}
||\Delta_A(l',t')||||\Delta_B(l'',t'')||\nonumber\\
=\frac12(||\Delta_A(l',t')||||\Delta_B(l'',t'')||+||\Delta_A(l',t')||||\Delta_B(l'',t'')||)\nonumber\\
\leq
||\Delta_A(l',t')||+||\Delta_B(l'',t'')||\nonumber
\end{eqnarray}
so that
\begin{equation}
|\langle O_A(t')O_B(t'')\rangle_C|\leq2[||\Delta_A(l',t')||+||\Delta_B(l'',t'')||]
+|\langle O_A^{l'}(t')O_B^{l''}(t'')\rangle_C|\nonumber
\end{equation}
Now we can substitute Eq.[\ref{corr01}] and Eq.[\ref{corr02}] in the inequality above and we obtain the following:
\begin{eqnarray}
&&|\langle O_A(t')O_B(t'')\rangle_C|\nonumber\\
&&\leq2||\Delta_A(l',t')||+2||\Delta_B(l'',t'')||+|\langle O_A^{l'}(t')O_B^{l''}(t'')\rangle_C|\nonumber\\
&&\leq2c\Big(|A|e^{-\frac{l'-v_{LR}|t'|}{a/2}}+|B|e^{-\frac{l''-v_{LR}|t''|}{a/2}}\Big)+\alpha e^{-\frac{d_{AB}-l'-l''}{\chi}}\nonumber
\end{eqnarray}
Finally we replace $l'$ and $l''$ with the optimal values $l'=\frac{\chi v_{LR}|t'|+\xi d_{AB}}{\chi+a}$ and $l''=\frac{\chi v_{LR}|t''|+\xi d_{AB}}{\chi+a}$ in order to obtain
\begin{equation}
|\langle O_A(t')O_B(t'')\rangle|_\psi\leq k e^{-\frac{d_{AB}}{\chi+a}}\left(e^{\frac{v_{LR}|t'|}{\chi+a}}+e^{\frac{v_{LR}|t''|}{\chi+a}}\right)^2
\end{equation}
}

We use this result to bound the quantum geometric tensor as this can be written as sum of unequal-time connected correlation functions.

 Let $\chi$ be the correlation length of $\ket{\psi_0(\lambda_0)}$. One has\cite{PhysRevLett.93.140402,LMS,2006CMaPh.265..119N}  $\chi=2{v_{LR,0}}{\Delta^{-1}}$  where $v_{LR,0}$ is the Lieb-Robinson speed associated to $H(\lambda)$. By  using  the clustering of correlations in Eq.(\ref{gcor})  one obtains
$
|q_1(t)|/N\leq k||\partial H^2||_M^2\sum_{j}\int_0^tdt'\int_0^tdt''\Big[e^{-\frac{d_{0j}}{\chi+a}}\Big(e^{\frac{v_{LR}|t'|}{\chi+a}}+e^{\frac{v_{LR}|t''|}{\chi+a}}\Big)^2 \Big]
$, 
where $||\partial H^q||_M^2$ is the maximum of the norms $||\partial H_j^q||^2$ and $d_{0j}$ is the Euclidean distance of the support of $\partial H^q_j$ from the origin.
A straightforward calculation leads to the following upper bound: 
\begin{eqnarray}\label{q_scaling}
|q_1(t)|/N&\leq&
k||\partial H^2||_M^2\Big[4t^2e^{\frac{2v_{LR}t}{\chi+a}}\sum_{j}e^{-\frac{d_{oj}}{\chi+a}} \Big]
\end{eqnarray}
To understand the behaviour of the above expression for large $N$, it is crucial to analyse the behaviour of the initial correlation length $\chi$. For a non critical Hamiltonian $H(\lambda)$, the gap is finite, thus the correlation length is finite  and the rescaled geometric tensor does not diverge for any $t$. Notice that this behaviour does not depend on the criticality of $H^q$. 
In view of the inequality $q(t)\leq q(0)+q_1(t)+2\sqrt{q(0)q_1(t)}$, we see that the divergences of $q(t)$ can only be those of $q(0)$. 
We therefore obtain the remarkable result that the phase diagram  on $\mathcal M_t$ is preserved in time.

\section{Equilibration of the quantum geometric tensor} 
In this section we want to show that the  geometric tensor, after the simplified quench, equilibrates, in the sense that its oscillations go to zero in the large $N$ limit. In  previous section we showed that for the simplified quench the quantum geometric tensor is  $q(t) =q_1(t)$  and that this can be written as sum of (connected) correlation functions $C(t',t'')$. 

We first prove that these correlation functions equilibrate. We will do so by showing that their temporal variance defined as $\sigma^2(C)=\lim_{T\rightarrow\infty}T^{-2}\int_0^Tdt'\int_0^Tdt''(C(t',t'')-\overline{C(t',t'')})^2$ is small. Indeed, since the probability for $C(t_1,t_2)=x$ 
is given by $p(x)=\lim_{T\rightarrow\infty}T^{-2}\int_0^T\int_0^Tdt'dt''\delta(C(t',t'')-x)$, the expectation value for $C$ over the whole time interval is $E(C)=\lim_{T\rightarrow\infty}T^{-2}\int_0^Tdt'\int_0^Tdt''C(t',t'')=\overline{C(t',t'')}$ and therefore a small variance $\sigma^2$ means that the probability of observing a correlation function with a value different from its average is small, in other words, $\forall\lambda>0$, the probability $p(|C(t',t'')-\overline {C(t',t'')}|>\sigma\lambda)\leq\lambda^{-2}$.

It is known that the temporal variance for the expectation value of observables evolving under the non-resonance condition of the Hamiltonian\cite{PhysRevLett.101.190403} is bounded as $\sigma_A^2\leq\|A\|^2Tr(\overline{\rho}^2)$ where $Tr(\overline{\rho}^2)$ is the purity of the completely dephased  state $\overline{\rho} = \sum_n P_n^q\rho P_n^q$ in the basis of the evolving Hamiltonian, here $H^q = \sum_n E^q_n P^q_n $. This result can be  extended to unequal-time correlation functions, by means of the following

\noindent {\bf Theorem.} Consider a Hamiltonian $H=\sum_n E_nP_n$ satisfying the non-resonance condition, that is, being non degenerate and also having non degenerate gaps: $E_n-E_m= E_k-E_l\Rightarrow n=k \wedge m=l$. Then, the temporal variance $\sigma^2(C)$ of unequal-time correlation functions $C(t',t'') =\langle A(t')A(t'')\rangle$ is upper bounded as $\sigma^2(C)\le \| A\|^4 \tr \overline{\rho}^2$. 


{
\emph{Proof.} Let $\langle A(t')A(t'')\rangle$ be a unequal-times correlation function. 
The average of the function respect to the two involved times is $
\overline{Tr(A(t)A(t')\rho)}=\sum_{n}A_{nn}^2\rho_{nn}$. In order to find the variance we need the square of the oscillations around this average, that is
\begin{eqnarray}
&&(Tr(A(t)A(t')\rho)-\overline{Tr(A(t)A(t')\rho)})^2\nonumber\\&&=\sum_{mnl}\sum_{rst}A_{mn}A_{nl}\rho_{lm}e^{it(E_m-E_n)}e^{it'(E_n-E_l)}\nonumber\\&&\times(1-\delta_{mn}\delta_{nl})A_{rs}A_{st}\rho_{tr}e^{it(E_r-E_s)}e^{it'(E_s-E_t)}(1-\delta_{rs}\delta_{st})\nonumber
\end{eqnarray}
Finally we consider the double-times average of the above equation, that is the variance:
\begin{eqnarray}
\sigma^2&=&\overline{(Tr(A(t)A(t')\rho)-\overline{Tr(A(t)A(t')\rho)})^2}\nonumber\\
&=&\sum_{mnl}\sum_{rst}A_{mn}A_{nl}A_{rs}A_{st}\rho_{lm}\rho_{tr}\delta_{ms}\delta_{nr}\delta_{nt}\delta_{ls}(1-\delta_{mn}\delta_{nl})(1-\delta_{rs}\delta_{st})\nonumber\\
&=&\sum_{m\neq n}|A_{mn}|^4\rho_{mm}\rho_{nn}\leq\sum_{mn}|A_{mn}|^2A_{nm}\rho_{mm}A_{mn}\rho_{nn}\leq \max_{mn}|A_{mn}|^2Tr(A\overline{\rho})^2\nonumber
\end{eqnarray}
At this point is useful to remark the following inequality:
\begin{equation}
\max_{mn}|A_{mn}|=\max_{mn}|\langle m|A|n\rangle|^2\leq \max_k||A|k\rangle||^2\leq||A||^2\nonumber
\end{equation}
fow which
\begin{eqnarray}
\sigma^2&\leq&||A||^2Tr(A\overline{\rho})^2\nonumber
\end{eqnarray}
where we have exploited the non-resonance condition. Moreover, given a spectral resolution $A=\sum_ia_i|a_i\rangle\langle a_i|$ for $A$, we can write
\begin{eqnarray}
Tr(A\overline\rho)^2&=&\sum_{ij}a_ia_j\langle a_i|\overline{\rho}|a_j\rangle\langle a_j|\overline{\rho}|a_i\rangle\leq \max_i a_i^2\sum_{ij}\langle a_i|\overline{\rho}|a_j\rangle\langle a_j|\overline{\rho}|a_i\rangle\nonumber\\&=&||A||^2Tr(\overline{\rho}^2)\nonumber
\end{eqnarray}
and thus
\begin{eqnarray}
\sigma^2&\leq&||A||^4Tr(\overline{\rho})^2
\end{eqnarray}
\emph{Q.E.D.}
}

As a  corollary,  the same bound holds also for connected correlation functions.

Let us now apply these results to the temporal variance for the geometric tensor. From Eq.(\ref{gcor}), we see that we can write $q_1(t)$ in the form $q(t)=\int_0^tdt'\int_0^tdt''f(t',t'')$. Then we have  $q(t)=t^2 \overline{f(t',t'')} + X(t)$ where $X(t) = \int_0^tdt'\int_0^tdt''[f(t',t'') - \overline{f(t',t'')}]$. Obviously, 
$
|X(t)|/t^2 \leq t^{-2} \int_0^tdt'\int_0^tdt''|f(t',t'') - \overline{f(t',t'')}|
\leq\sqrt{t^{-2} \int_0^tdt'\int_0^tdt''[f(t',t'') - \overline{f(t',t'')}]^2}=\sigma 
$ and therefore $|X(t)|^2\leq t^2\sigma$. 
By applying the result of the above theorem, the long-time behaviour of the geometric tensor is thus given by
\begin{eqnarray}\label{bound01}
q_1(t)&=&t^2\overline{\langle \partial H^q(t')\partial H^q(t'')\rangle_C}+X\nonumber\\
|X|&\leq& t^2 |\partial H^q|^4Tr(\overline{|\psi_0\rangle\langle\psi_0|}^2)
\end{eqnarray}

We can now state one of the main results of the paper. Let us consider a Hilbert space with $N$ bodies, such that its dimension behaves according to  $d=q^N$. If the initial state $\ket{\psi_0}$ is sufficiently spread in the eigen-basis of $H^q$, that is, its purity is proportional to $1/d$,  then,  time fluctuations, $X$, are upper bounded as $|X|\le q^{-N} |\partial H^q|^4 = q^{-N} O(N^4)$ so that they vanish very fast for a large system.

In the case of the general quench, one has to make use of the triangular inequality $q(t)=q(0)+q_1(t)+|Y|$ where $|Y|\leq2\sqrt{q(0)q_1(t)}$.
{
By exploiting the bound in Eq.(\ref{bound01}),  this expression becomes
\begin{equation}
q(t)=q(0)+t^2\overline{\langle \partial H^q(t')\partial H^q(t'')\rangle_C}+(X+Y)
\end{equation}
with
\begin{eqnarray}
|X+Y|&\leq&|X|+ 2\sqrt{q(0)}\sqrt{t^2\overline{\langle \partial H^q(t')\partial H^q(t'')\rangle_C}+X}\nonumber\\
&\leq&2t\sqrt{q(0)\overline{\langle \partial H^q(t')\partial H^q(t'')\rangle_C}}+|X|+2\sqrt{q(0)|X|}\nonumber
\end{eqnarray}
where in the last line we have used $\sqrt{a+b}\leq\sqrt{a}+\sqrt{b}$.
}

Therefore we can state that as for large $N$ the purity suppresses $|X|$,  the QGT $q(t)$ oscillates around $q(0)+t^2\overline{\langle \partial H^q(t')\partial H^q(t'')\rangle_C}$ at most linearly in time.

We  conclude this section with a remark.
{
 In the simplified quench protocol, the geometric tensor $q(t)$ can be expressed as in Eq.(\ref{gt}) with $\partial H=0$. If we define $\Delta=E_1-E_0$ and exploit the resolution of the identity $\sum_{n}\ket{\psi_n}\bra{\psi_n}$ then we can also see that $|q(t)|\le \Delta^{-2}\sum_{n\neq0}|\langle\psi_0|[H,D]|\psi_n\rangle|^2 = \Delta^{-2}\langle[H,D][H,D]^\dag\rangle $. By making explicit the expression of $H$ and $D$ as functions of the local interactions $H_i$ we finally obtain
\be
|q(t)|\le\frac{1}{\Delta^2}\sum_{ijkl}\int_0^tdt'\int_0^tdt''\langle[H_i,\partial H^q_j(t')][H_k,\partial H^q_l(t'')]^\dag\rangle
\ee
}

The eigenvalues of the geometric tensor are thus upper bounded  by the sum of OTOCs. This means that regions of higher distinguishability correspond to the large operator spreading of the local terms in the Hamiltonian. Moreover, one can see that the time fluctuations of the quantum geometric tensor  are directly connected to the time fluctuations of the OTOCs, and this, in turn,  provides a framework to  unify different aspects of quantum dynamics like quantum chaos and scrambling\cite{otocANDchaos,otoc2,otocANDscr}. Recently, it has been shown that in the quantum Dicke model a {notion of  out-of-time-order fidelity}  is connected to both entanglement spreading and chaos\cite{fotoc}. In our approach, the appearance of OTOCs is a generic feature of the space-time description for the geometry of quantum states. 

\begin{figure}
  \centering
  \includegraphics[scale=.5]{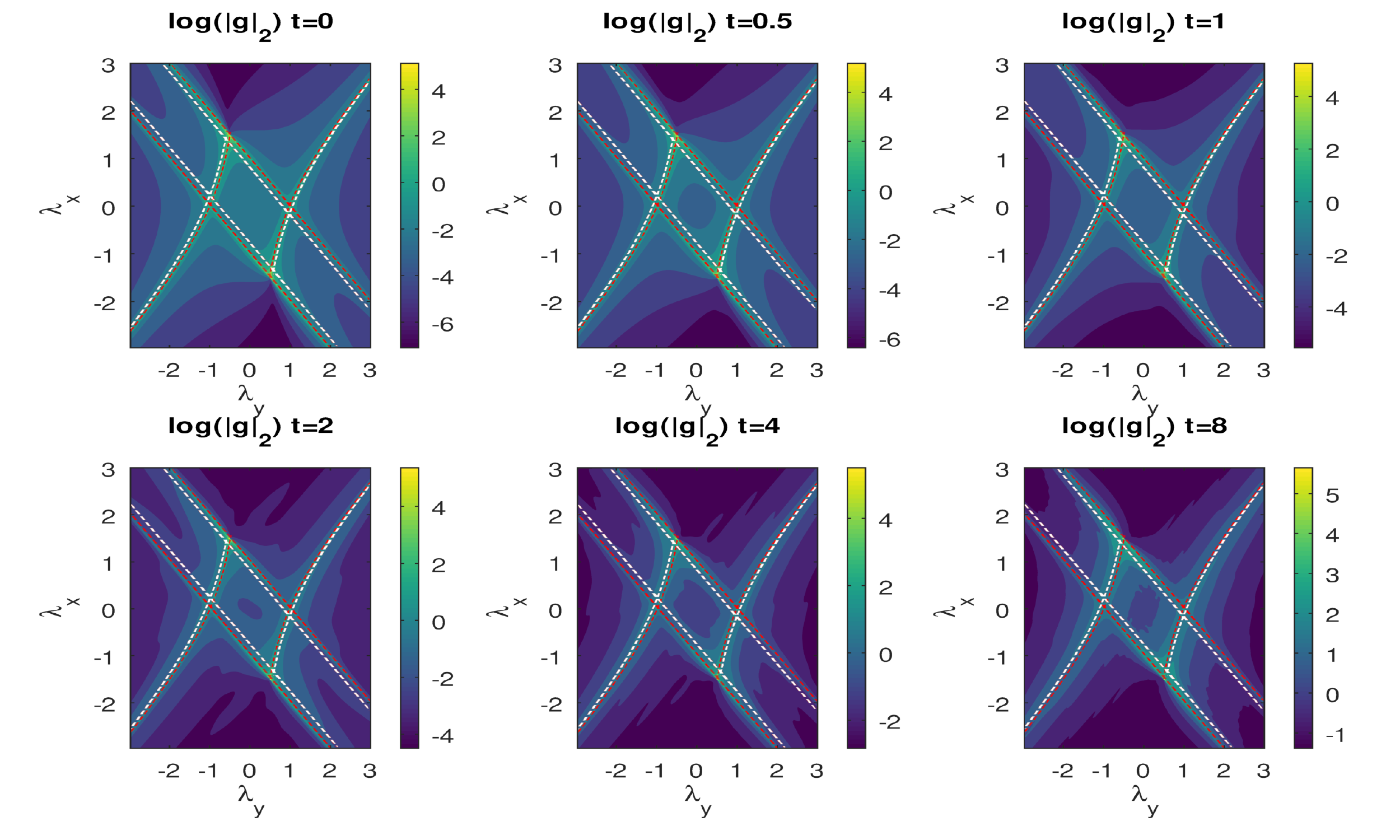}
  \caption{Time evolution of the logarithm of the norm of the rescaled metric $g_{\mu\nu}(t)/N$ after the orthogonal quench $(\lambda_x,\lambda_y,h=0)\mapsto (\lambda_x,\lambda_y,h=0.2)$ for $N=500$ spin. The red dashed lines represent the critical lines at $t=0$. The white dashed lines represent the states under a critical quench. We see that the singularities of the metric tensor only depend on those at $t=0$. }
  \label{fig1}
\end{figure}
%

\section{QGT in the Cluster-XY model} 
We now apply these findings in an exactly solvable spin chain. In\cite{operatorqgt}, the QGT was promoted from states to operators, and that treatment bears some similarity with ours, especially in its application to spin chains.
In order to demonstrate meaningfully the time evolution of the geometric tensor after the orthogonal quench, we need a model with at least three parameters. We consider the Cluster-XY model\cite{hammaclusterxy}. The model interpolates between a stabiliser Hamiltonian and the quantum XY model. The stabiliser Hamiltonian is the sum of  terms of the form $K_\mu = \sig{\mu}{z}\prod_{\nu\sim\mu}\sig{\nu}{x}$ where $\mu,\nu$ label the sites of a lattice and $\nu\sim\mu$ denotes that $\nu$ is connected to $\mu$. The ground state for this Hamiltonian is important as it is a universal resource for measurement based quantum computation\cite{MBQC-Original, MBQC-Nature}.
 The Hamiltonian reads
\ba
 H=&-\sum_{i=1}^N \sig{i-1}{x}\sig{i}{z}\sig{i+1}{x}-h\sum_{i=1}^N\sig{i}{z}&+\la_y\sum_{i=1}^N \sig{i}{y}\sig{i+1}{y}+\la_x\sum_{i=1}^N \sig{i}{x}\sig{i+1}{x}
\label{clusterXY}
\ea
A somehow canonical way to prepare the quench is the following. Let us indicate  the manifold parameters with  $\lambda=(\lambda', \bar{\lambda})$ and consider the  submanifold $\mathcal M_{\bar{\lambda}}$ where the parameters $\bar{\lambda}$ are   fixed. The {\em orthogonal} quench is given by $H(\lambda',\bar{\lambda})\mapsto H^q \equiv H(\lambda',\bar{\lambda}+q)$. This sudden quench produces a time evolution on the quantum geometric tensor on $\mathcal M_{\bar{\lambda}t}$.

The Hamiltonian Eq.(\ref{clusterXY})  can be diagonalized by the standard technique of Jordan-Wigner transformation $c_l^\dagger=\left(\prod_{m=1}^{l-1}\sig{m}{z}\right)\sig{l}{+}$
that maps the model  to a quadratic Hamiltonian of spinless fermions $\{c_n,c_m\}=0,\, \{c_n,c_m^\dagger\}=\delta_{nm}$, followed by Fourier transform, $c_k=\frac{1}{\sqrt N}\sum_{n=1}^N e^{ikn}c_n,\quad k=\frac{\pi}{N}(2m+1)$, after which the Hamiltonian reads  
\begin{equation*}
 H=2\sum_{0\leq k\leq\pi}\left[\epsilon_k(c_k^\dagger c_k+c_{-k}^\dagger c_{-k})+i\delta_k(c_k^\dagger c_{-k}^\dagger+c_k c_{-k})\right]
\end{equation*}
Finally, a Bogoliubov transformation diagonalizes the above Hamiltonian in each $k$ block by $ \gamma_k=\cos(\theta_k)c_k-i\sin(\theta_k)c_{-k}^\dagger$\cite{hammaclusterxy},  where $\theta_k=-1/2\arctan{\delta_k}/{\epsilon_k}$ with $\delta_k=\sin(2k) - (\lambda_x - \lambda_y)\sin(k)$ and $\epsilon_k=\cos(2k) - (\lambda_x + \lambda_y)\cos(k) - h$. 
\begin{figure}
  \centering
  \includegraphics[scale=.5]{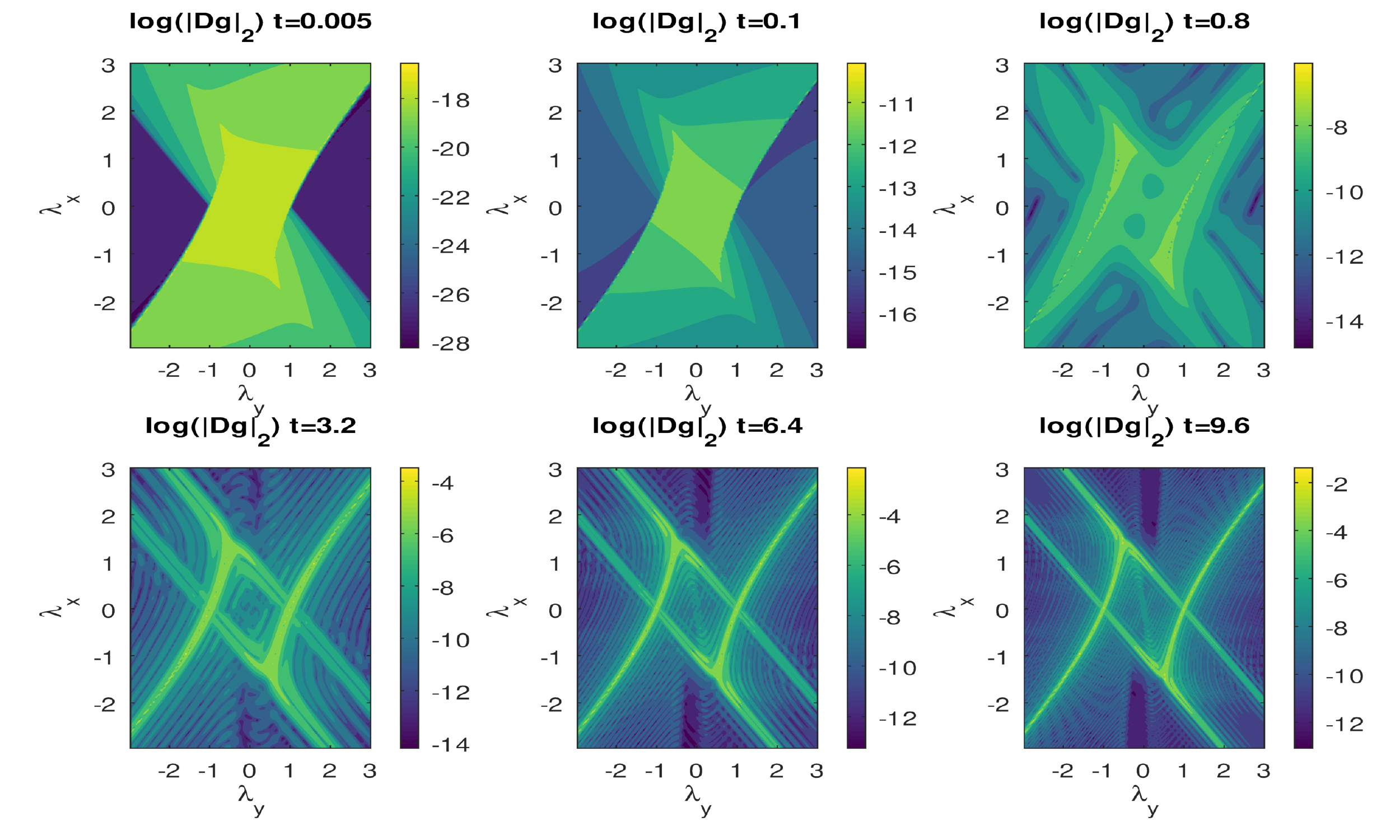}
  \caption{Time evolution of $\log | \Delta g_{\mu\nu} (t)/N|$ after a small orthogonal quench $(\lambda_x,\lambda_y,h=0)\mapsto (\lambda_x,\lambda_y, h=0.001)$ for $N=500$ spins. Starting from a completely zero metric at $t=0$, the time dependent part starts developing lines  higher values for the modulus of $| \Delta g_{\mu\nu} (t)/N|$ that correspond to the critical lines of the initial Hamiltonian. These lines, though, do not correspond to real divergences of the geometric tensor but to regions of higher distinguishability. }
  \label{fig2}
\end{figure}
%
This allows to write the time evolution after a quantum quench in an exact way.

{
Let $|\Omega(\lambda)\rangle$ be the ground state of $H(\lambda)$. Its time evolution by the quench Hamiltonian $H(\lambda+q)$ has been calculated in\cite{hammaclusterxy} and it  is given by:
\begin{equation}\label{evo01}
|\Omega(\lambda,t)\rangle=\prod_{0\le k\le\pi}\Big[\cos(\chi_k)+i e^{-4it\Delta_k(\lambda+q)}\sin(\chi_k)\gamma_k^\dag(\lambda+q)\gamma_{-k}^\dag(\lambda+q)\Big]|\Omega(\lambda+q)\rangle
\end{equation}
where $\Delta_k(\lambda)=\sqrt{\delta_k^2+\epsilon_k^2}$ is the energy of a Bogoliubov particle of momentum $k$ and $\chi_k=\theta_k(\lambda)-\theta_k(\lambda+q)$.
For our purposes it is easier to work with an expression of the evolved state as a function of the fermionic operators  $c_k^\dag$, $c_k$ that are independent on  $\lambda$. The ground state then reads\cite{hammaclusterxy}:
\begin{equation}\label{eq:XY_grounda}
|\Omega(\lambda)\rangle=\prod_{0\le k\le\pi}\Big(\cos(\theta_k(\lambda))+i\sin(\theta_k(\lambda))c_k^\dag c_{-k}^\dag\Big)|0\rangle_c\nonumber
\end{equation}
where $|0\rangle_c$ is the vacuum state for the $c_k$ operators. Substituting in Eq.\ref{evo01} the expression of the $c_k$ operators in therms of  $\gamma_k$ and exploiting their fermionic algebra, one obtains
\ba\nonumber
|\Omega(\lambda,\!t)\!\rangle\!=\!\!\!\!
\prod_{0\le k\le\pi}\!\!\!\Big\{\!\!\cos(\chi_k)[\cos(\theta_k) \!+ \!i \sin(\theta_k)c_k^\dag c_{-k}^\dag]+i e^{-4it\Delta_k}\sin(\chi_k)[\cos(\theta_k)c_k^\dag c_{-k}^\dag + i\sin(\theta_k)]\Big \}|0\rangle
\ea
where the functions $\theta_k$ and $\Delta_k$ are calculated in $\lambda+q$.

At this point one can compute directly the quantum geometric tensor from the squared fidelity $\mathcal F^2$:
\begin{eqnarray}
\mathcal F^2&\equiv&|\langle\Omega (\lambda',t)|\Omega (\lambda,t)\rangle\nonumber|^2\\
&=&\prod_{0\le k\le\pi}\Big|\cos(\theta_k'-\theta_k)\cos(\chi_k')\cos(\chi_k)+ \cos(\theta_k'-\theta_k)e^{4it(\Delta_k'-\Delta_k)}\sin(\chi_k')\sin(\chi_k) \nonumber\\&&+ \sin(\theta_k'-\theta_k)e^{-4it\Delta_k}\cos(\chi_k')\sin(\chi_k)-\sin(\theta_k'-\theta_k)e^{4it\Delta_k'}\sin(\chi_k')\cos(\chi_k)\Big|^2\nonumber
\end{eqnarray}\
where
\begin{eqnarray}
\Delta_k &=& \Delta_k(\lambda + q)\nonumber\\
\Delta_k' &=& \Delta_k(\lambda+\delta\lambda+q)\nonumber\\
\chi_k &=& \theta_k(\lambda) - \theta_k(\lambda + q)\nonumber\\
\theta_k'-\theta_k&=&\theta_k(\lambda+\delta\lambda+q)-\theta_k(\lambda + q)\nonumber\\
\chi_k' &=& \theta_k(\lambda+\delta\lambda) - \theta_k(\lambda+\delta\lambda+q).\nonumber
\end{eqnarray}
Since the Hamiltonian is real, the quantum geometric tensor is real and it is thus a Riemannian metric. To obtain it, we look at the fidelity in the second order for the infinitesimal shift, $\delta\lambda$. 
We obtain
\ba\label{eq.guvXY0}
g_{\mu\nu} (t) = g_{\mu\nu} (0) + \Delta g_{\mu\nu} (t)\nonumber
\ea
where $g_{\mu\nu} (0) = N^{-1}\sum_{0\le k\le\pi}\partial_\mu\theta_k\partial_\nu\theta_k$ and the time dependent term is given by
\begin{eqnarray}\label{eq.guvXY}
\Delta g_{\mu\nu} (t)&=&\frac 1N\sum_{0\le k\le\pi}\Big\{\partial_\mu\theta_k'\partial_\nu\theta_k' (2-2\cos(4t\Delta_k')\nonumber\\
&-&4\sin^2(4t\Delta_k') \cos^2(\chi_k)\sin^2(\chi_k))+[\partial_\nu\theta_k'\partial_\mu\theta_k + \partial_\mu\theta_k'\partial_\nu\theta_k](\cos(4t\Delta_k')-1)\nonumber\\
&-& 4t\sin(4t\Delta_k')[\partial_\mu\theta_k'\partial_\nu\Delta_k'+ \partial_\nu\theta_k'\partial_\mu\Delta_k']\cos(\chi_k)\sin(\chi_k)[1-2\sin^2(\chi_k)]\nonumber\\
&+& 16t^2\sin^2(\chi_k)(1-\sin^2(\chi_k))\partial_\mu\Delta_k'\partial_\nu\Delta_k')\Big\}
\end{eqnarray}
where $\theta_k=\theta_k(\lambda)$, $\theta_k'=\theta_k(\lambda + q)$, $\chi_k = \theta_k - \theta_k'$ and $\Delta_k'=\Delta_k(\lambda + q)$.
}
By analyzing the latter it is possible to acquire all information concerning   the time evolution the metric tensor, including the time evolution of the phase diagram and its equilibration. Let us first show that the phase diagram is conserved, that is, no new critical lines are added on top of the ones at $t=0$, nor the original ones are deformed.

{
By inspection of Eq.(\ref{eq.guvXY}) we see that divergences can appear only in the terms $\partial_\mu\theta_k$, $\partial_\mu\theta'_k$ and $\partial_\mu\Delta'_k$,} and this may only happen when $\Delta_k$ (or $\Delta_k'$) becomes null for some $k\in[-\pi,\pi)$ in the thermodynamic limit. Since these gaps appear in different terms, the phase diagram is not deformed. It must be the one at time zero, plus possibly the phase diagram of the quench Hamiltonian.

{
Let us show that the phase diagram of the quench Hamiltonian does not add any critical line to the phase diagram of the evolving system defined by $g_{\mu\nu}(t)$. This happens when in the thermodynamic limit the function $|g_{\mu\nu}(t)-g_{\mu\nu}(0)|$ is bounded. We start with 
\begin{equation}
|g_{\mu\nu}(t)-g_{\mu\nu}(0)|\leq A+B+4tC+16t^2D\nonumber
\end{equation}
where
\begin{eqnarray}
A&=&\max_k2|\partial_\mu\theta_k'\partial_\nu\theta_k'(1-\cos(4t\Delta_k')-2\sin^2(4t\Delta_k') \cos^2(\chi_k)\sin^2(\chi_k))|\nonumber\\
B&=&\max_k |\Big[\partial_\nu\theta_k'\partial_\mu\theta_k + \partial_\mu\theta_k'\partial_\nu\theta_k\Big](\cos(4t\Delta_k')-1)|\nonumber\\
C&=&\max_k |\sin(4t\Delta_k')[\partial_\mu\theta_k'\partial_\nu\Delta_k' + \partial_\nu\theta_k'\partial_\mu\Delta_k']\sin(2\chi_k)[\frac12-\sin^2(\chi_k)]|\nonumber\\
D&=&\max_k |\sin^2(\chi_k)(1-\sin^2(\chi_k))\partial_\mu\Delta_k'\partial_\nu\Delta_k'|\nonumber
\end{eqnarray}
and then we consider
\begin{eqnarray}
\partial_\mu\theta_k'&=&-\frac12\frac{\epsilon_k'\partial_\mu\delta_k'-\delta_k'\partial_\mu\epsilon_k'}{\Delta_k'^2}\;\;\;{\rm and} \nonumber \\ \partial_\mu\Delta'_k&=&\frac{\epsilon_k'\partial_\mu\epsilon'_k+\delta_k'\partial_\mu\delta'_k}{\Delta'_k}.\nonumber
\end{eqnarray}
 Since both $\partial_\mu\delta_k'$ and $\partial_\mu\epsilon_k'$ are bounded functions, the following bounds hold
\begin{eqnarray}
|\partial_\mu\theta_k'|&\leq&\frac12\Big[\Big|\frac{\epsilon_k'\partial_\mu\delta_k'}{\sqrt{\epsilon_k'^2+\delta_k'^2}}\Big|+\Big|\frac{\delta_k'\partial_\mu\epsilon_k'}{\sqrt{\epsilon_k'^2+\delta_k'^2}}\Big|\Big]\frac{1}{\Delta_k'}\leq F_\mu \frac{1}{\Delta_k'}\nonumber\\
|\partial_\mu\Delta'_k|&\leq&\Big|\frac{\epsilon_k'\partial_\mu\delta_k'}{\sqrt{\delta_k'^2+\delta_k'^2}}\Big|+\Big|\frac{\epsilon_k'\partial_\mu\epsilon_k'}{\sqrt{\epsilon_k'^2+\delta_k'^2}}\Big|\leq G_\mu
\end{eqnarray}
where $F_\mu$ e $G_\mu$ are not diverging. Then only $\partial_\mu\theta_k'$ may diverge at most as $\Delta_k'^{-1}$. Thus,  it is immediate to see that all divergences in $A$, $B$, $C$ and $D$ are canceled by multiplication with terms that  go to zero at least linearly with  $\Delta_k'$.
}
The result is that as $\Delta_k'\rightarrow 0$ no new divergence in $g_{\mu\nu} (t)$ is introduced: the phase diagram is conserved by the temporal evolution. 

In Fig.\ref{fig1} we plot the time evolution of $\log |g_{\mu\nu} (t)|$ in the plane $\lambda_x,\lambda_y$ after the orthogonal quench $h=0\mapsto h=0.2$.  The initial phase diagram is clearly visible. We have superimposed the lines corresponding to the criticality of the quenching Hamiltonian. As the time evolution proceeds, the initial divergences stay constant but they are thickened as regions of higher distinguishability (which is also, larger curvature). However, this higher curvature never diverges. In other words, the phase diagram is constant. In Fig.\ref{fig2} we plot just the logarithm of the norm of the rescaled $\Delta g$, that is, the time dependent part. 

In order to show the equilibration properties, we  first analytically compute the purity of the dephased state $\overline{\rho}$,
{
where $\rho=|\Omega(\lambda,t)\rangle\langle\Omega(\lambda,t)|$.  If we call $|c_n|^2$ the populations of the state $\Omega(\lambda)$ in the eigenbasis of $H(\lambda+q)$, this purity reads $\tr\overline{|\Omega\rangle\langle\Omega|}^2=\sum_n|c_n|^4$. At this point from Eq.(\ref{evo01}) one easily obtains that} $Tr(\overline{\rho}^2)= \prod_k[\cos(\chi_k)^4+\sin(\chi_k)^4]= \prod_k(1-1/2\sin^2(2\chi_k))$, which shows that for a quench $\chi_k = \theta_k-\theta'_k$ the purity of the dephased state $\bar{\rho}$ is exponentially small in $N$ and therefore, in view of the theorem, equilibration follows.

\section{Conclusions and Outlook} 
In this paper, we have shown that the metric structure that describes the geometry of the manifold of ground states of a family of Hamiltonians can be extended to non-equilibrium states, for example states that evolve unitarily under a quantum quench. We have shown that the initial phase diagram is conserved and that the geometric tensor equilibrates. 
One of the interesting aspects  of this formulation is that the geometric tensor can be written in terms of out-of-time-order commutators. This suggests that the study of the fluctuations of the geometric tensor can be useful to investigate questions of quantum chaos and  the transition to non-integrability. Moreover, the connection of a space-time metrics with OTOCs can be important in order to understand scrambling in black holes\cite{scrINbh}; for instance, one would be interested in knowing what kind of geometric structure corresponds to a fast scrambler and use these insights to reconstruct Hamiltonian models for a black hole. It would be interesting to show how the fluctuations of the QGT are connected to spreading and complexity of entanglement\cite{ess, yoshida} - and their relation to the many-body localisation transition - or  to investigate  the emergence of irreversibility in quantum mechanics\cite{irr}. In this respect, we would like to understand the Riemannian curvature as a probe for a transition in different entanglement spectrum statistics and dynamical behavior (integrable, ergodic, localized). As an unrelated problem, it would be important to use these methods to address questions in Adiabatic Quantum computation,  as its performance depends on the curvature along adiabatic evolution\cite{brac}. One could then perturb the Hamiltonian in order to obtain shortcuts to adiabaticity by flattening the curvature along the adiabatic evolution\cite{delcampo}. In perspective, the study of the space-time behaviour of the quantum geometric tensor may provide an unifying framework for the study of quantum dynamics.


\end{document}